\documentclass[12pt]{iopart}

\usepackage{caption}
\usepackage{subcaption}
\usepackage{graphicx}
\usepackage[symbol]{footmisc}

\begin{document}

\title{Telescope imaging beyond the Rayleigh limit in extremely low SNR}

\author{Hyunsoo Choi\footnote[2]{The authors contributed equally to this work.}, Seungman Choi\footnotemark[2], Peter Menart\footnotemark[2], Angshuman Deka, and Zubin Jacob$^{*}$}

\ead{zjacob@purdue.edu} 
\address{Birck Nanotechnology Center, School of Electrical and Computer Engineering, Purdue University, West Lafayette, IN 47907, USA}

\vspace{10pt}
\begin{indented}
\item[]January 2024
\end{indented}

\begin{abstract}

The Rayleigh limit and low Signal-to-Noise Ratio (SNR) scenarios pose significant limitations to optical imaging systems used in remote sensing, infrared thermal imaging, and space domain awareness. In this study, we introduce a  Stochastic Sub-Rayleigh Imaging (SSRI) algorithm to localize point objects and estimate their positions, brightnesses, and number in low SNR conditions, even below the Rayleigh limit. Our algorithm adopts a maximum likelihood approach and exploits the Poisson distribution of incoming photons to overcome the Rayleigh limit in low SNR conditions. In our experimental validation, which closely mirrors practical scenarios, we focus on conditions with closely spaced sources within the sub-Rayleigh limit (0.49-1.00R) and weak signals (SNR less than 1.2). We use the Jaccard index and Jaccard efficiency as a figure of merit to quantify imaging performance in the sub-Rayleigh region. Our approach consistently outperforms established algorithms such as Richardson-Lucy and CLEAN by 4X in the low SNR, sub-Rayleigh regime. Our SSRI algorithm allows existing telescope-based optical/infrared imaging systems to overcome the extreme limit of sub-Rayleigh, low SNR source distributions, potentially impacting a wide range of fields, including passive thermal imaging, remote sensing, and space domain awareness. 

\end{abstract}

\vspace{2pc}

% \submitto{\NJP}

\section{Introduction}
Traditional super-resolution techniques were developed for the field of microscopy where the sample/object to be imaged is accessible \cite{Schermelleh2019SR}. Thus, imaging resolution can be enhanced by labeling with unique light emitters \cite{BLom2017STED, Kianinia2018Quantumemitters, Schwartz2021SR_Quantum}, using patterned illumination sources, or engineering nano-scatterers in proximity to the object of interest \cite{Heintzmann2017SRM} (e.g., metamaterials near the object \cite{Lee2021MetaNanoscopy, choi2023fluorescence}). In stark contrast, for remote sensing \cite{Wang2019RemoteSensing}, thermal imaging \cite{Rivadeneira2020ThermalIS}, and space domain awareness \cite{Holzinger2020SDA} applications, the object of interest is inaccessible due to large distances. Thus, super-resolution techniques in telescopes have received less attention since the object cannot be labeled or illuminated, nor can there be engineered proximal scatterers. 

In the case of such passive remote sensing, unique challenges also exist due to the low signal-to-noise ratio (SNR) regime. In many situations, long exposure times are not possible, and thus the SNR has to be sacrificed to accommodate this constraint \cite{tandra2008snr}. Additionally, images of distant real-world objects through a telescope are cluttered by background noise as they are not in a controlled lab environment. Therefore, low SNR scenarios have long been studied, including techniques for improving sensor capabilities \cite{mandracchia2020fast}, building robustness in low SNR \cite{coupe2012candle}, and recovering images from low SNR images \cite{guan2022deep}. In addition, deconvolution \cite{richardson1972bayesian,fish1995blind,preibisch2014efficient,yang2020deconvolution} and denoising methods \cite{buades2005review,fan2019brief,tian2020deep} have proven to be effective in dealing with low SNR conditions. The goal of this paper is to put forth an approach for existing telescopic imaging systems to function in a new extreme regime of low SNR and sub-Rayleigh source distributions with minimal or no apriori information. 

The Rayleigh criterion describes the best resolution of a telescope based on classical optics. Sources with a separation smaller than the Rayleigh limit are considered irresolvable, meaning it is not possible to determine the unknown number of sources or their respective positions and brightnesses \cite{Zhou2019BrightnessEstimate}. The need for accurate estimation of object (label-free light source) characteristics like location and relative brightness remains crucial in multiple applications, such as space domain awareness \cite{bao2021quantum}, remote sensing \cite{long2017accurate}, and infrared thermal imaging. Due to the prevalence of this Rayleigh constraint, numerous research initiatives have sought to surmount this challenge \cite{frohn2000true,ram2006beyond,tamburini2006overcoming}. Recently, the integration of quantum imaging techniques offers the prospect of surpassing the diffraction limit \cite{paris2009quantum,tsang2016quantum}; however, their practical implementation necessitates intricate imaging setups or complex devices, such as spatial light modulators (SLMs) or digital micromirror chips (DMDs) \cite{tang2016fault,smith2017single}. Additionally, many studies have reported results obtained only under controlled laboratory conditions, such as with a predefined number of sources, equal brightness sources, or the absence of background noise, which plays a crucial role in estimation \cite{kurdzialek2023measurement}. Consequently, these outcomes do not adequately address practical, real-world scenarios where delicate tuning may be infeasible and acquiring any prior information is impossible. We also note that antibunching photon statistics from point dipolar emitters is often used as a quantum resource for super-resolution \cite{Schwartz2021SR_Quantum}. However, our work is focused on the fields of remote sensing, thermal imaging, and space domain awareness, where the extended light sources exhibit bunched thermal photon statistics.

To address these practical concerns, we propose a Stochastic Sub-Rayleigh Imaging (SSRI) algorithm that leverages conventional imaging devices to resolve multiple sources in low SNR conditions. SSRI calculates the likelihood for different parameter sets based on the captured intensity image and maximizes this likelihood to estimate the parameters of interest (figure \ref{fig1}(a)). In this paper, we demonstrate SSRI's capacity to correctly estimate scenes with multiple sources that have different brightnesses and sub-diffraction limit separation, even with an exceptionally low SNR of less than 1.2. Notably, no known prior study has attempted to implement algorithms within such a markedly poor SNR setting, where visual perception fails to decipher individual sources or even differentiate between the presence and absence of signals amidst elevated noise levels. To test the efficacy of SSRI, we applied the algorithm to experimental data in a variety of SNRs, separation levels, and relative source brightness conditions. To quantify imaging resolution in the sub-Rayleigh regime, we use the metrics of Jaccard index and Jaccard efficiency. We show that our approach consistently outperforms well-established state-of-the-art algorithms such as Richardson-Lucy and CLEAN.

Along with improved estimation of the number of sources, SSRI also demonstrates the ability to accurately estimate source location for smaller separations than the Richardson-Lucy and CLEAN algorithms. SSRI correctly determines the source position for separations as low as half the Rayleigh limit, compared to a minimum separation of 0.86 times the Rayleigh limit for Richardson-Lucy or CLEAN. This means that by incorporating SSRI into a telescope system with a 4 m aperture and operating at 500 nm wavelength, we can pinpoint objects less than 0.08 meters apart at 1000 km away ($<$ 0.53R). This marks a 38\% or more improvement in resolution over traditional methods, which can only discern objects at a minimum spacing of 0.13 meters (0.86R) under similar conditions.

We believe our algorithm can be used as a new benchmark to evaluate the performance of any future quantum \cite{Tan2021SR_SPADE, Tsang2019SR_SPADE} or classical \cite{Zhang2015ClassicalRayleigh} superresolution techniques. Recent advances have primarily focused on improving quantum superresolution \cite{Tan2021SR_SPADE} without detailed comparison with highly optimized direct imaging methods. As a result, our proposed algorithm serves as a simple yet highly effective benchmark to guide future research in this area. In this paper, we shall expound upon the underpinning theory of our algorithm, demonstrate its practical implementation, and assess its effectiveness in coping with an array of challenging test situations.

\section{SSRI Algorithm}

To estimate the positions, brightnesses, and number of point sources in a measured image, we implement a maximum likelihood parameter estimation method which we term Stochastic Sub-Rayleigh Imaging (SSRI). For the likelihood calculation, for a given certain measurement, (1) derive the equation that governs the intensity profile of a computed scene for a certain number of sources and each position and brightness. (2) Calculate the likelihood of measuring the scene for a certain parameter set inspired by the photon statistics. (3) Following the maximum likelihood estimation approach, find the proper parameter set. 
For $N$ number of sources with centroid coordinates $(x_n, y_n)$ and brightness $b_n$ for the n-th star, the scene intensity $I(x,y)$ is given by
\begin{equation}
    I(x, y) = \sum_{n=1}^{n=N} b_n P(x-x_n, y-y_n)
\end{equation}
where $P$ is the point-spread function (PSF) determined by the aperture of the imaging system and $x$ and $y$ are image plane coordinates. The probability $p$ of measuring a particular number of photons $h_{x,y}$ at a camera pixel, assuming a mean photon rate $H_{x,y}$, is then given by a Poisson distribution

\begin{equation}
    p(h_{x,y}) = \frac{{H_{x,y}}^{h_{x,y}} e^{-H_{x,y}}}{h_{x,y}!}
\end{equation}
where, $I_m$ is the measured image, and $h_{x,y}$ be the photons measured at each camera pixel for this measured image. For each camera pixel with coordinates $(x, y)$, we expected $H_{x,y}$ photons to arrive at the camera pixel based on the scene intensity distribution $I(x,y)$ above. 
The photon counts at each pixel are identically, independently distributed variables, so we can calculate the probability of the measured image by multiplying the probabilities for each pixel. This results in the probability of the measured image for the unknown parameter set $\Theta= \{N,x_n,y_n,b_n \vert n=1,\cdots,N \}$, which is the likelihood $L$ that we desire. Therefore, the likelihood is

\begin{equation} \label{eq3}
L(\Theta \vert I_m) = \prod_{x,y}p(h_{x,y})
\end{equation}

where the product is taken over all pixels. The set of parameters that maximizes this likelihood is taken as the estimated scene. Details on the algorithm implementation can be found in the Materials and Methods section.

\section{Experimental Results}

\begin{figure} [hbt!]
\includegraphics[width=\linewidth]{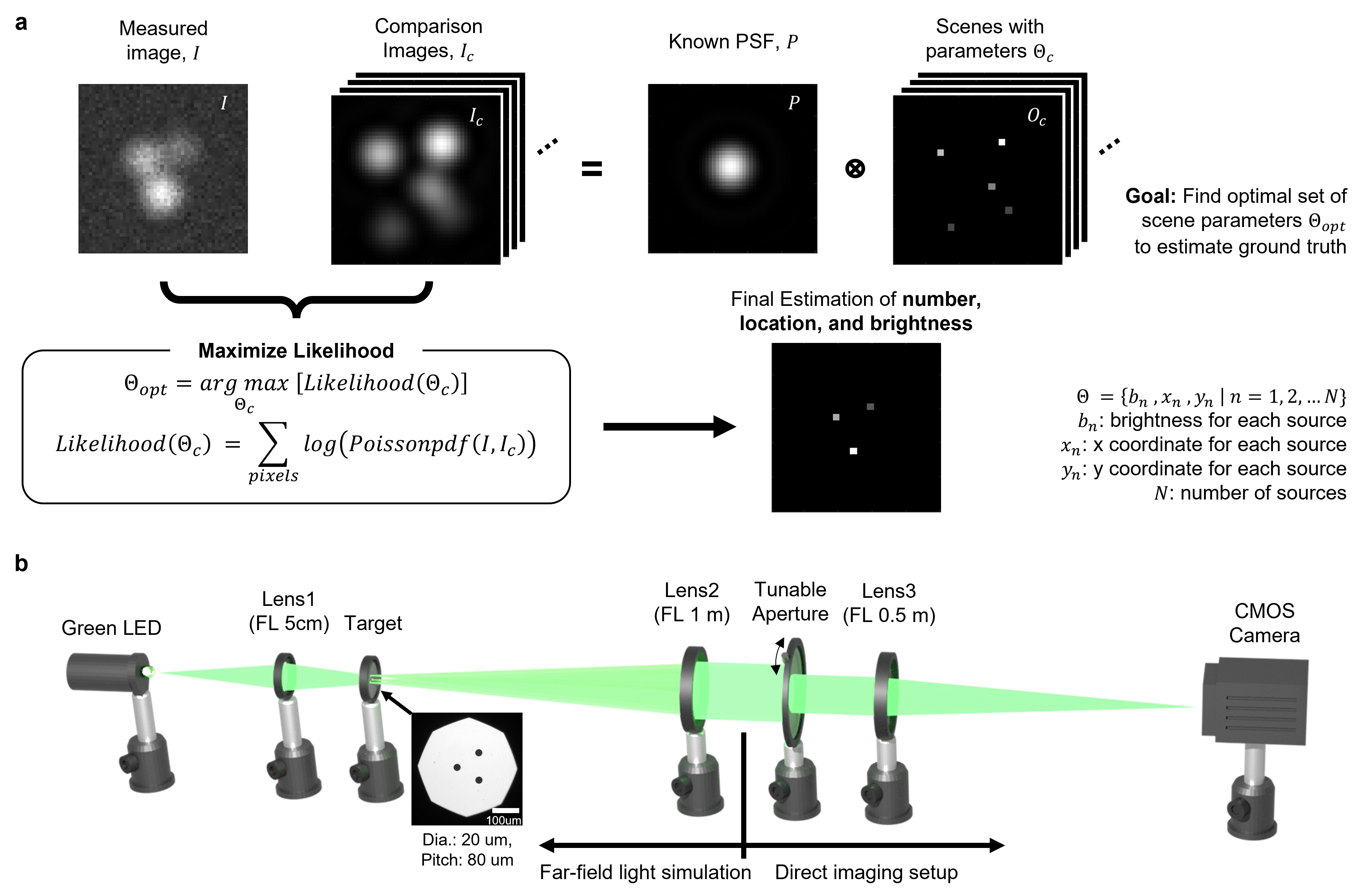}
\caption{SSRI Algorithm and Experiment. \textbf{a} Diagram illustrating the SSRI algorithm. Given source positions and brightnesses, an expected image can be generated using the PSF of the imaging system and is compared to the measured image. To estimate the ground truth scene, the source positions and brightnesses that maximize the likelihood are found.
\textbf{b} Experimental setup used to collect images for SSRI testing. Inset: Microscopic image of three pinholes fabricated by lithography technique (diameter: 20 $\mathrm{\mu m}$, distance between each pinhole: 80 $\mathrm{\mu m}$). } 
\label{fig1}
\end{figure}

The experimental setup is shown in figure \ref{fig1}(b). The light source is a green LED (M530L4, Thorlabs), which is focused on 20 $\mathrm{\mu m}$ diameter holes fabricated with a lithographic technique. A lens is placed after the targets to collimate the light from the pinholes. The lens is placed a long distance (1 m) from the targets, ensuring a flat wavefront and simulating light from a faraway point source. An iris and an additional lens are used as a telescope, and a complementary metal-oxide semiconductor (CMOS) camera (Atlas 2.8 megapixel, Lucid Vision) is used to collect images. The size of the aperture (iris) is varied to change the size of the PSF and obtain different ratios of target separation to PSF radius. Ground truth positions of the targets are determined using a large aperture size, resulting in a small PSF and easily distinguishable sources. Experiments are then conducted with a small aperture to obtain separations below the Rayleigh limit. The relative brightness of the targets is controlled by adjusting the position of the targets near the edge of the illuminated region, resulting in a different level of illumination for each pinhole.  

\begin{figure} [hbt!]
\includegraphics[width=\linewidth]{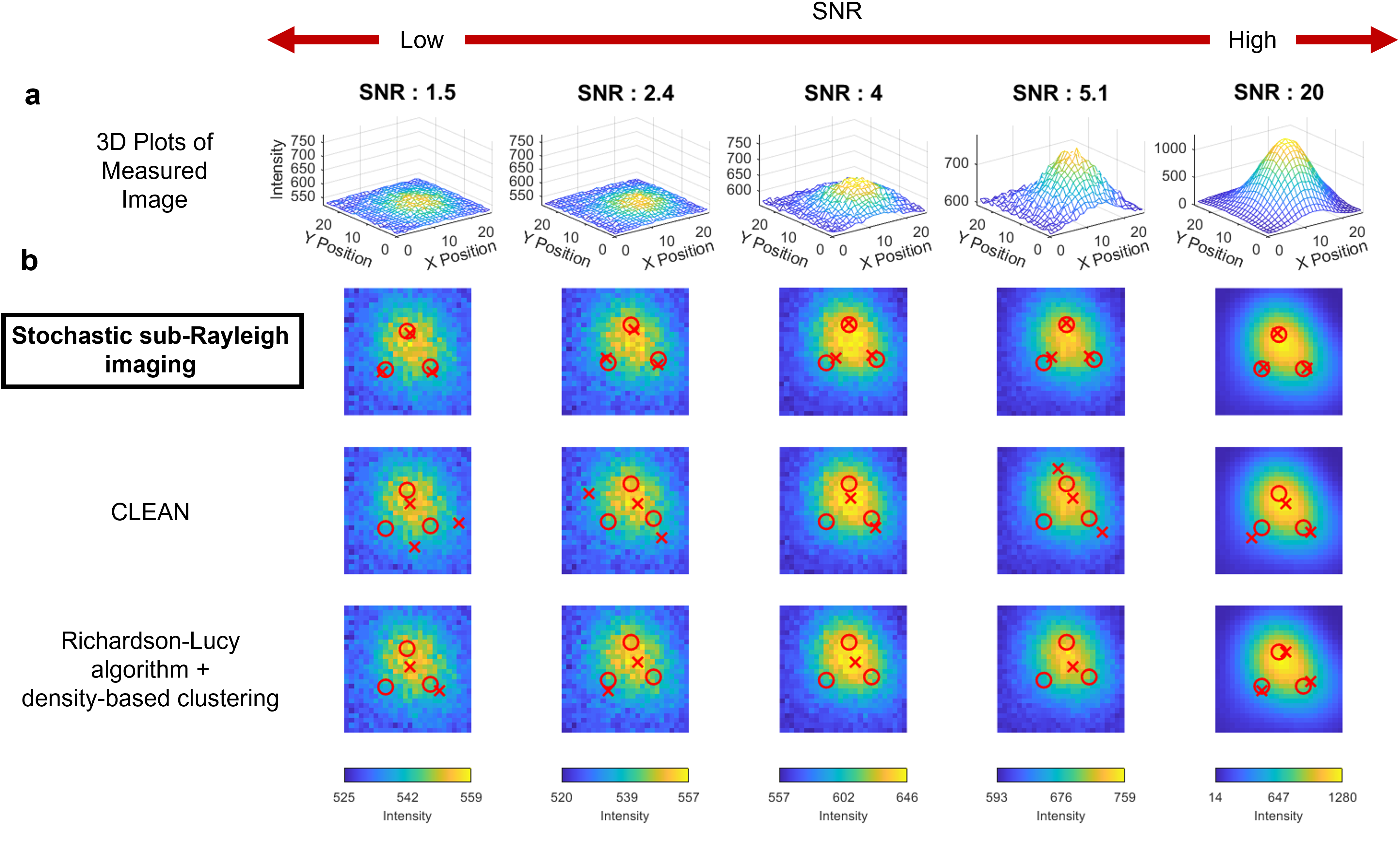}
\caption{Low SNR position and number estimation using SSRI, CLEAN, and Richardson-Lucy algorithms. \textbf{a} 3D mesh plots of the measured images to illustrate the difference in the size of the signal versus fluctuations from the noise for different SNRs. \textbf{b} Estimation results for low SNR scenes and one high SNR scene. The image consists of three equidistant point sources arranged in a triangular array with 0.6 Rayleigh limit separation between the sources. The relative brightness values of the point sources are [1, 2, 3] counterclockwise from the bottom left. SSRI shows the best estimation result for low SNR conditions. All algorithms correctly estimate the source number in high SNR.} 
\label{fig2}
\end{figure}

\subsection{Low SNR Position and Number Estimation}

Low SNR is a challenge faced in many imaging applications. Our focus in this article is on telescopic systems engaged in remote sensing, infrared thermal imaging and space domain awareness. However, this problem is of interest even in biomedical imaging. For example, particle tracking is a bioimaging technique where nanoscopic reporter particles are designed to attach to biomolecules of interest. The optical signal from these point-like reporter particles is then used to track the dynamics of the biomolecules. Low signal occurs in situations where low exposure times are used to track fast-moving molecules, or when the reporter particles do not emit a large number of photons, \cite{manza2015sptreview} resulting in SNRs below 4 \cite{chenouard2014sptcomp}. In addition, it is often necessary to use a low density of reporter particles to avoid overlap of PSFs from different reporters 
\cite{manza2015sptreview}. 

Given the importance of low SNR imaging, we tested the SSRI method's performance in a variety of low SNR conditions. This is shown in figure \ref{fig2}, along with results from two highly developed deconvolution algorithms, Richardson-Lucy and CLEAN, for comparison. 
Using all three methods, we also attempted to estimate the number of sources and their positions in a three source scene with ~0.6 times the Rayleigh limit separation between sources. Figure \ref{fig2}(a) shows 3D plots of the images to more clearly illustrate the size of the signal versus the size of the noise fluctuations in the low SNR conditions. Figure \ref{fig2}(b) shows the results of the estimations from each of the three algorithms. Note that a density-based clustering method is used after deconvolution with the Richardson-Lucy algorithm. Refer to the Materials and Methods section for details on the implementation of the Richardson-Lucy and CLEAN algorithms, the SNR calculation, and the quantifying metric calculation.

We observe that while Richardson-Lucy and CLEAN algorithms correctly estimate the source number in the high SNR condition, they fail to correctly estimate both the number of sources and source positions in low SNR conditions, while SSRI correctly estimates 3 sources in both the high and low SNR images. SSRI also estimates the source positions with only small position errors, less than 2-3 pixels, in all cases. Note that air movement causes the target location to move by up to 2 pixels, meaning it is difficult to determine the ground truth location with better accuracy than 2 pixels. 

We also quantified the comprehensive imaging performance of those methods in the sub-Rayleigh region using the Jaccard index and Jaccard efficiency (See more detail in Materials and Methods section) \cite{sage2019super}. In Table \ref{tab1}, SSRI shows a great margin of advance in Jaccard efficiency, consistently showing 4X better performance especially in low SNR. Overall, SSRI demonstrates superior performance in the sub-Rayleigh, low SNR region compared to Richardson-Lucy and CLEAN, and good performance even for an SNR of 1.5.

\begin{table}[hbt!]
\begin{tabular}{cccccccccccc}
                           & \multicolumn{5}{l}{Jaccard Index} &  & \multicolumn{5}{l}{Jaccard Efficiency}        \\
SNR                        & 1.5   & 2.4  & 4    & 5.1  & 20   &  & 1.5   & 2.4   & 4     & 5.1   & 20    \\ \cline{1-6} \cline{8-12} 
\multicolumn{1}{c|}{SSRI}  & 100   & 100  & 100  & 100  & 100  &  & 98.75 & 98.47 & 98.13 & 98.37 & 99.19 \\
\multicolumn{1}{c|}{CLEAN} & 0     & 0    & 25   & 0    & 50   &  & 0     & 0     & 24.87 & 0     & 49.88 \\
\multicolumn{1}{c|}{RL}    & 25    & 25   & 0    & 0    & 100  &  & 24.81 & 24.78 & 0     & 0     & 97.93
\end{tabular}
    \caption{Table for Jaccard index and Jaccard efficiency for various SNRs showing SSRI is the only method to have near 100\% efficiency in low SNR regime}
\label{tab1}
\end{table}

\begin{figure} [hbt!]
\includegraphics[width=\linewidth]{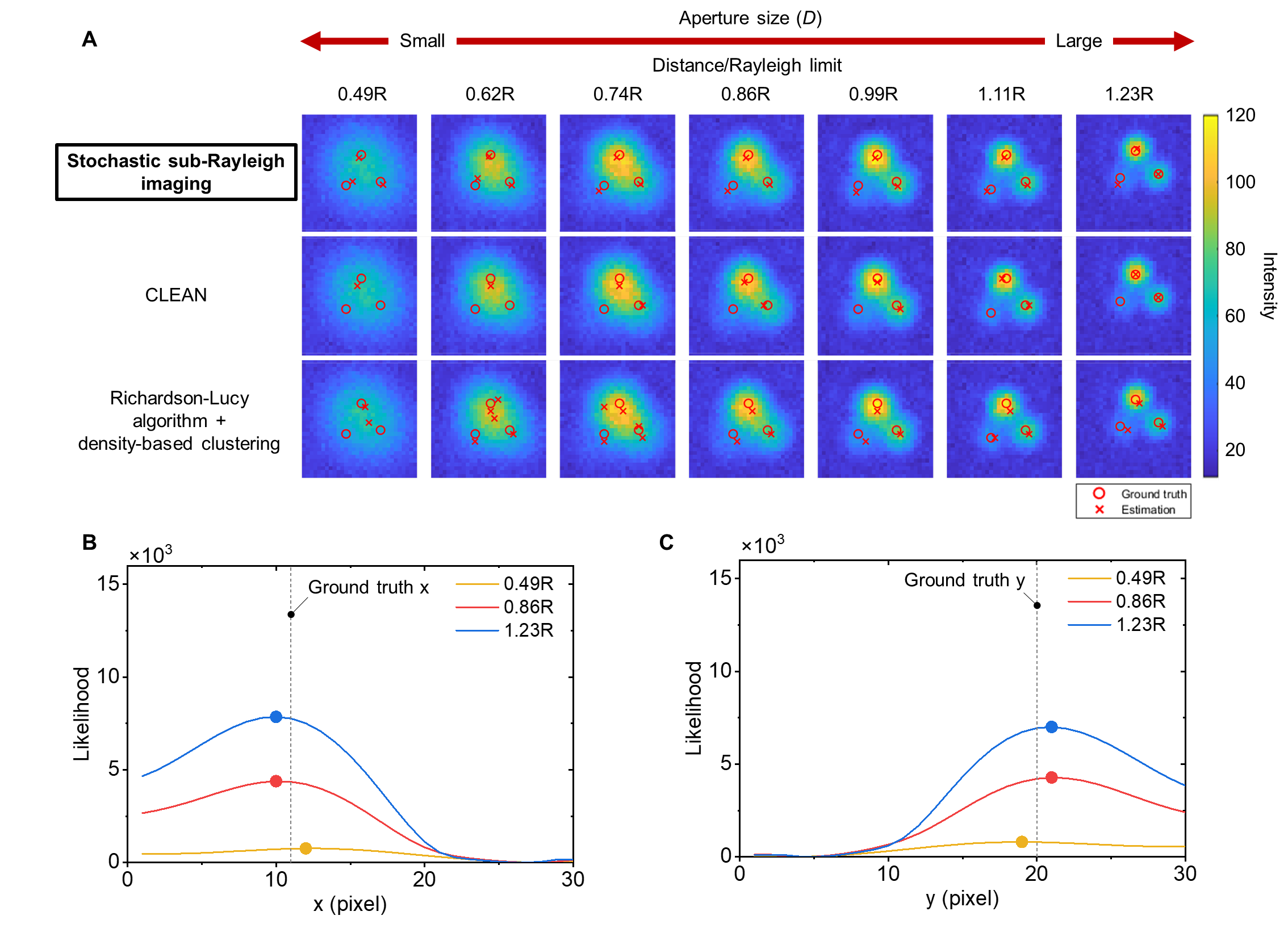}
\caption{Centroid estimation for varying aperture size. (A) Centroid estimation results using stochastic sub-Rayleigh imaging (SSRI), CLEAN, and Richardson-Lucy deconvolution. The image consists of three equidistant point sources arranged in a triangular array with a distance of 80 $\mathrm{\mu m}$ between them. The relative brightness values of the point sources are [1, 2, 3] counterclockwise from the bottom left. The signal-to-noise ratio (SNR) is set to approximately 7 for a point source with a relative brightness of 1. At all aperture sizes, SSRI outperforms CLEAN and Richardson-Lucy with accurate centroid estimation. (B-C) Likelihood analysis for a single parameter, specifically the x and y-coordinate center of the point source with a relative brightness of 1. The likelihood is plotted against different aperture sizes, demonstrating the relationship between aperture size and parameter sensitivity. The position of the point source at the largest likelihood is highlighted with a dot.} 
\label{fig3}
\end{figure}  

\begin{table}[]
    \begin{subtable}[]{0.5\linewidth}
        \centering
        \begin{tabular}{cccccccc}
                                   & \multicolumn{5}{l}{Jaccard Index}     & \multicolumn{1}{l}{} &       \\
        SNR                        & 0.49R & 0.62R & 0.74R & 0.86R & 0.99R & 1.11R                & 1.23R \\ \hline
        \multicolumn{1}{c|}{SSRI}  & 100   & 100   & 100   & 100   & 100   & 100                  & 100   \\
        \multicolumn{1}{c|}{CLEAN} & 33.33 & 33.33 & 66.67 & 66.67 & 66.67 & 66.67                & 66.67 \\
        \multicolumn{1}{c|}{RL}    & 25    & 60   & 60   & 100   & 50    & 100                  & 100  
        \end{tabular}
    \end{subtable}

    \begin{subtable}[]{0.5\linewidth}
        \begin{tabular}{cccccccc}
                                   & \multicolumn{5}{l}{Jaccard Efficiency}        &       &       \\
        SNR                        & 0.49R & 0.62R & 0.74R & 0.86R & 0.99R & 1.11R & 1.23R \\ \hline
        \multicolumn{1}{c|}{SSRI}  & 92.11 & 92.93 & 92.68 & 92.62 & 91.99  & 94.15 & 94.98  \\
        \multicolumn{1}{c|}{CLEAN} & 29.68  & 30.40  & 65.02 & 65.59 & 65.59 & 65.92 & 66.67 \\
        \multicolumn{1}{c|}{RL}    & 22.93 & 58.67 & 58.75 & 88.67 & 48.58 & 91.05 & 90.25
        \end{tabular}
    \end{subtable}
    \caption{Table for Jaccard index and Jaccard efficiency for various aperture sizes showing SSRI has up to 3.97 times better efficiency than other state-of-the-art algorithms}
    \label{tab2}
\end{table}

\subsection{Various Rayleigh Ratio Estimation}

The diffraction limit in space imaging is a significant constraint on angular spatial resolution, marking a key hurdle for scientists. Nevertheless, our approach has surpassed this standard resolution, achieving an angular resolution of more than twice that of traditional telescopes. To test SSRI's performance for sub-diffraction limit separations, we varied the aperture size to change the ratio of the target separation and the diffraction-limited resolution as dictated by the Rayleigh limit. Figure \ref{fig3}(a) illustrates the centroid estimation performance at different aperture sizes using our SSRI method, CLEAN, and Richardson-Lucy algorithms along with the Jaccard index and efficiency presented in table \ref{tab2}. The image comprises three equidistant point sources arranged in a triangular array, with a distance of 80 $\mathrm{\mu m}$ between them. The relative brightness ratio of the point sources is adjusted as [1, 2, 3] counterclockwise from the bottom left.

Throughout the aperture size range, CLEAN fails to estimate the correct number of point sources by neglecting the weakest point source. It estimates 2-point sources in the range of 0.74-1.23R, while below 0.74R, it identifies only one source. In addition, CLEAN tends to yield more position estimation error, which results in a drop in the Jaccard efficiency. On the other hand, the Richardson-Lucy algorithm successfully estimates 3-point sources when the aperture size is large enough, corresponding to separations from 0.86-1.23R. However, as the aperture size becomes smaller and the separation reaches 0.74R, it tends to inaccurately estimate the number of point sources and their locations yielding false positive (FP) detection, which results in a decrease in the Jaccard index and efficiency. In contrast, our SSRI method consistently provides correct estimates for the number of point sources across all aperture ranges, with relatively small centroid estimation errors below 2 pixels. Overall, SSRI shows more than 90\% efficiency whereas CLEAN and RL show below 70\% efficiency, especially in a sub-Rayleigh region (below 0.74R).

Figure \ref{fig3}(b) and (c) present the likelihood function of SSRI within a single parameter space for scenes with varying aperture sizes. The parameters of interest are the x and y-coordinates of a point source with a relative brightness of 1 (the bottom-left point source in figure \ref{fig3}(a)), while the other parameters remain fixed. For all aperture sizes, the likelihood functions take the form of a parabolic curve with the maximum value lying around the ground truth, albeit with varying slopes. As the aperture size decreases, the slope around the apex becomes less steep, resulting in reduced sensitivity to parameter changes. Consequently, the maximum likelihood centroid position may slightly vary with the aperture size. Nevertheless, even at the smallest aperture size (4 mm), the SSRI algorithm demonstrates reliable performance with errors below 2 pixels. These findings underscore the ability of SSRI to overcome the blurring effect in low SNR and consistently provide accurate estimates for the number and position of point sources.

\subsection{Brightness Estimation} 

The accurate measurement of both the location and brightness of point light sources is vital in imaging/tracking resident space objects. In the case of fluorescence-based bio-imaging \cite{lifante2020role, Anila2023ChemSocRev}, the blurring effect induced by fluorescence poses significant challenges, specifically in achieving precise brightness estimation and localizing fluorescent tissues accurately. Overcoming these obstacles and obtaining reliable brightness estimation hold the key to unlocking new possibilities in biomedical imaging, enhancing overall accuracy, and advancing clinical applications. In this section, we focus on the ability of SSRI to estimate source brightness.

\begin{figure} [hbt!]
\includegraphics[width=\linewidth]{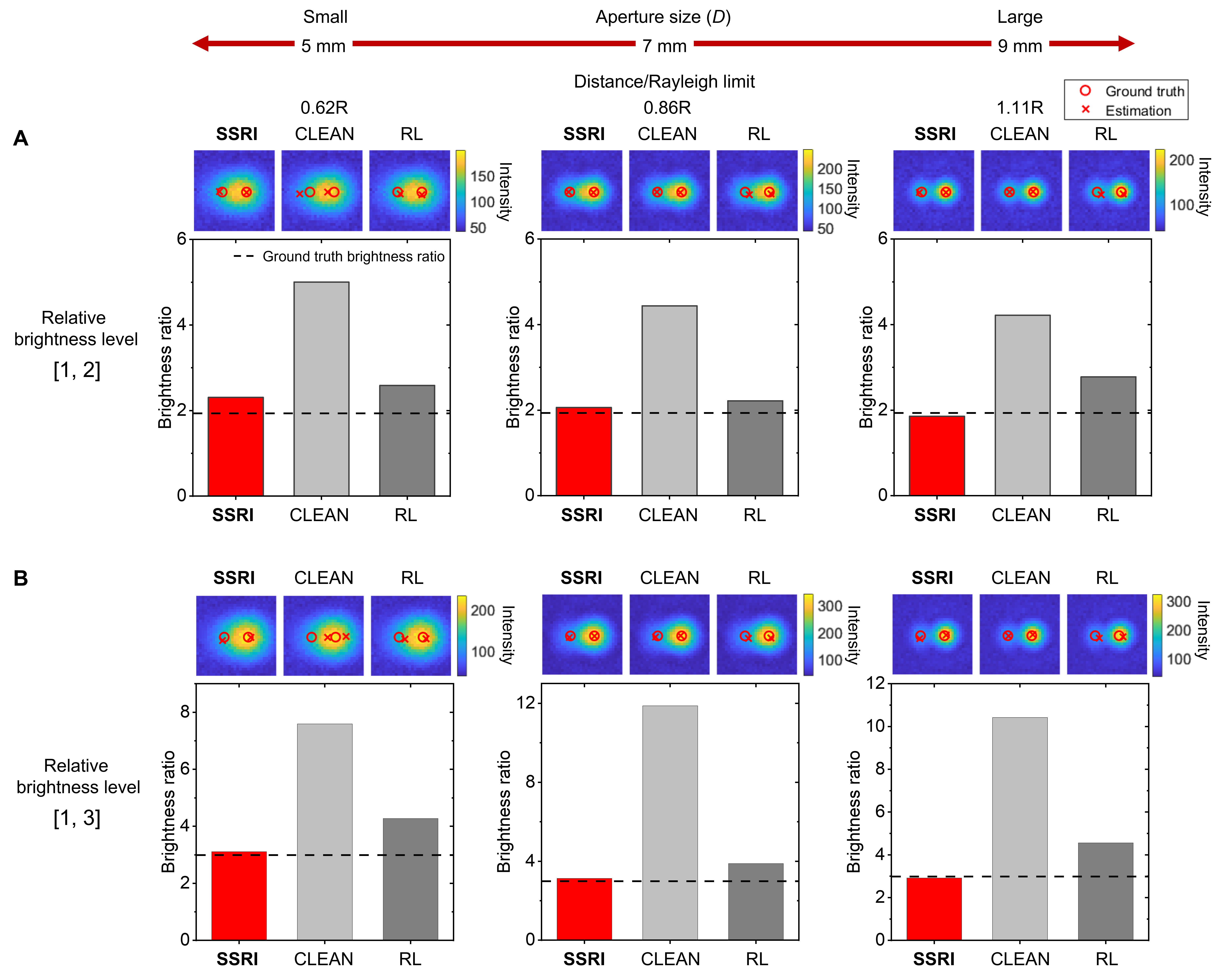}
\caption{Brightness estimation comparison using stochastic sub-Rayleigh imaging (SSRI), CLEAN, and Richardson-Lucy deconvolution. Aperture size vs. centroid and brightness estimation results for a scene consisting of 2-point sources with relative brightness of [1,2] (A) and [1,3] (B). point sources are spaced apart at 80 $\mathrm{\mu m}$ intervals and SNR is set to around 13 for all the cases. At all aperture sizes, SSRI outperforms CLEAN and RIcharson-Lucy with accurate bright estimation.} 
\label{fig4}
\end{figure}

Figure \ref{fig4} illustrates position and brightness estimation results for 2-point sources obtained using the SSRI, CLEAN, and Richardson-Lucy deconvolution methods. The experimental setup involved laterally displacing two pinhole targets with a separation of 80 $\mathrm{\mu m}$ to achieve varying relative brightness levels: [1,2] (figure \ref{fig4}(a)) and [1,3] (figure \ref{fig4}(b)). Measurements were conducted at three different aperture sizes (5, 7, and 9 mm). To enable a focused comparison of relative brightness estimation, K-means clustering was applied for CLEAN and Richardson-Lucy, leveraging prior knowledge of two-point sources only in the comparison group algorithms. 

Both CLEAN and Richardson-Lucy deconvolution methods exhibited precise centroid estimation outcomes mainly due to the incorporation of prior information regarding the number of point sources, with the exception observed in CLEAN's performance with a separation level of 0.62R. Meanwhile, the SSRI method achieved accurate centroid estimation performance for all cases despite not using any prior information.

Additionally, SSRI successfully estimated the correct brightness ratio for all cases, with a margin of error of approximately 0.2. In contrast, CLEAN and Richardson-Lucy deconvolution methods failed to accurately estimate brightness, often yielding higher relative brightness ratios compared to the ground truth. This discrepancy can be attributed to these methods' tendency to assign greater weight to intense point sources while neglecting signals from weaker point sources when multiple point sources coexist within a scene. We conclude that CLEAN and Richardson-Lucy deconvolution methods are insufficient for accurately resolving scenes with multiple point sources exhibiting varying levels of brightness within the sub-Rayleigh limit. On the other hand, the SSRI method demonstrated an accurate estimation of source brightness, number, and position in such complex scenarios.

\begin{figure} [hbt!]
\includegraphics[width=\linewidth]{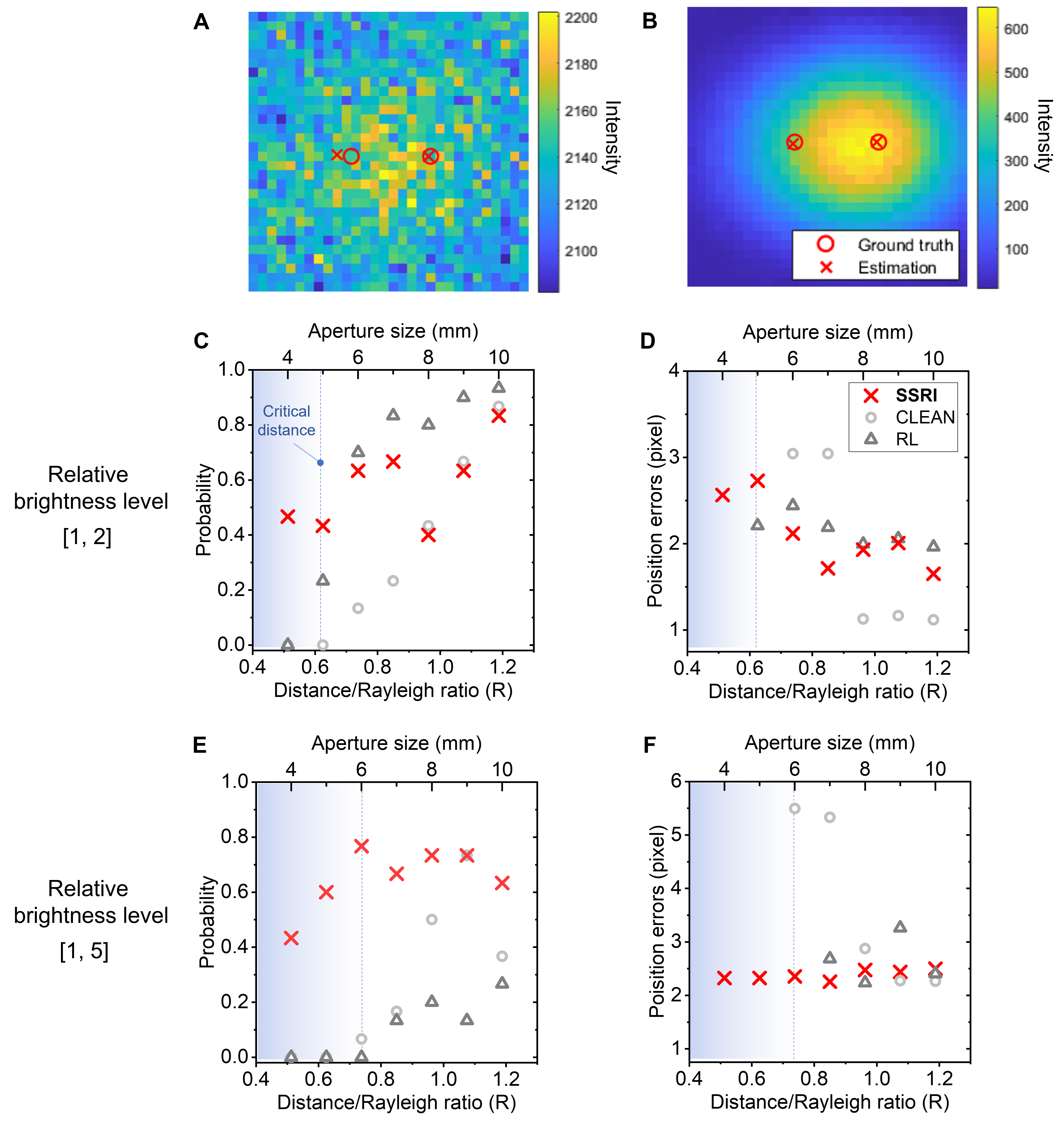}
\caption{SSRI estimation statistics for low SNR $<$ 1.2. Centroid estimation results for two-point source images of SNR 1 (a) and 20 (b) with relative brightness of [1,5]. Probability of correct number estimation and corresponding centroid estimation error in the relative brightness of [1,2] (c, d) and [1,5] (e, f). These probabilities are from 30 different images for each condition. SSRI provides an accurate estimation compared to the control group, especially for separations smaller than the Rayleigh criterion.} 
\label{fig5}
\end{figure} 

\subsection{Monte Carlo Simulation for extreme low SNR}

SSRI's performance was evaluated under challenging conditions involving extremely low SNR and a large relative brightness ratio of [1,5] (figure \ref{fig5}(a) and (b)). Remarkably, even under the extremely low SNR condition where visual perception cannot discern the existence of point sources due to severe noise, SSRI estimated centroids well. Further statistical analysis was conducted for 30 different scenes captured with SNR $<$ 1.2 and relative brightness ratios of [1, 2] and [1, 5], which is shown in figure \ref{fig5}(c)-(f). For scenes with distances between two-point sources larger than the Rayleigh limit (0.99-1.23R), both CLEAN and Richardson-Lucy algorithms correctly estimated the number of point sources with high probabilities ($>$0.4) and small position errors ($<$2 pixels). However, their correct number estimation probabilities decreased as the distance decreased, failing completely below the critical distance of 0.62R and 0.74R for relative brightness ratios of [1,2] and [1,5], respectively.

In contrast, SSRI showed less dependence on distance, consistently achieving correct number estimation probabilities over 0.4 even below the critical distance, with centroid errors below 3 pixels as shown in figure \ref{fig5}(c) and (d). This trend was more evident in scenes with a large relative brightness ratio of [1, 5] (figure \ref{fig5}(e) and (f)), where SSRI outperformed CLEAN and Richardson-Lucy in both probability and position error across separation conditions from 0.49-0.86R. SSRI's correct number estimation probability remained above 0.4 in all separation conditions, with small position errors below 2.5 pixels, probably given by the uncertainty in ground truth position due to air movement.

These results demonstrate SSRI's reduced dependence on distance and stable performance under extremely low SNR conditions. In contrast, the CLEAN and Richardson-Lucy algorithms showed increasing errors as the distance decreased for both brightness ratios. These results highlight SSRI's superior centroid estimation performance in challenging sub-Rayleigh imaging scenarios with extremely low SNR and large relative brightness ratios, surpassing CLEAN and Richardson-Lucy deconvolution methods.

In summary, these results demonstrate, for the first time, breaking the Rayleigh limit in a very low SNR scenario without prior information using only a conventional telescopic imaging device. This can also serve as a baseline comparison method for further research in classical/quantum super-resolution imaging.

\section{Conclusion}

In conclusion, our novel SSRI algorithm presents an advance in addressing the longstanding challenge of overcoming the Rayleigh limit in telescope imaging systems, particularly under low SNR conditions. We experimentally demonstrated that SSRI consistently surpasses established algorithms like Richardson-Lucy and CLEAN in accurately determining the position, brightness, and count of point sources under different separation and SNR scenarios.
We note that SSRI could serve as an effective benchmark for assessing the performance of quantum superresolution techniques, as it addresses conventional imaging challenges that have not been directly confronted in the quantum domain thus far. This development can help guide future research efforts in imaging and localization methods, ultimately leading to more accurate solutions for a broad spectrum of telescope imaging applications.

\section{Materials and Method}
\subsection{SSRI Algorithm}

To perform the estimation, an upper limit on the number of point sources $N_\textnormal{max}$ is chosen. We choose $N_\textnormal{max} = 5$ for all results in this paper. The likelihood function in Eq.(\ref{eq3}) is maximized at $N=N_\textnormal{max}$ to find the brightnesses and source positions that maximize the likelihood and these values are taken as the estimated image. Then, sources outside of the region of interest or below a relative brightness threshold are discarded to achieve the final estimated image. For results in this paper, we discard sources within one pixel of the edge of the image and use a relative brightness threshold of 0.1.

Note that evaluating the likelihood function at different values $N$ to find the number of sources that maximize the likelihood is not a suitable approach to estimate the source number. The largest number of sources will give the largest likelihood regardless of the measured image, as additional sources provide more degrees of freedom. For example, it is generally possible to improve the likelihood by adding very dim sources that fit the noise. Thus, it is best to simply maximize the likelihood for the largest number of sources to be estimated, and then apply criteria to remove extraneous sources.

\subsection{CLEAN Algorithm}

The CLEAN deconvolution algorithm enhances radio astronomy images by iteratively identifying and subtracting point sources from a "dirty" image convolved with the given PSF, $P(x,y)$. This process is represented by the equation: 
\begin{equation}
R(x, y) = D(x, y) - aP(x - x_p, y - y_p), 
\end{equation}

where $D(x, y)$ is the dirty image, $R(x, y)$ is the residual image, $a$ is the scaling factor, and $(x_p, y_p)$ is the peak intensity point. The algorithm begins by locating the peak intensity point in the dirty image, corresponding to the brightest source. It then subtracts a scaled version of the PSF centered on this point, effectively removing the contribution of the detected source. The subtracted value is saved, and the process is repeated, identifying and subtracting additional point sources from the residual image. The iterative cleaning continues until a stopping criterion is met or the residual image reaches a specified threshold, ensuring significant sources are accurately estimated. The final clustering happens after the iterative process is done. Here, with the stored peak position sets $(x_p,y_p)$, using density-based clustering in the paper, we can correctly estimate the centroid of point sources. By iteratively cleaning the image in this manner, the algorithm estimates the centroid of the point sources, removing the effect of blur from the diffraction imposed by finite aperture.

\subsection{Richardson-Lucy algorithm}

The Richardson-Lucy algorithm is an iterative image deconvolution method used to restore degraded images. Given an observed or degraded image, $D(x, y)$, which is the result of convolving the true image, $I(x, y)$, with the PSF, represented as $P(x, y)$:
\begin{equation}
D(x, y) = I(x, y) * P(x, y)
\end{equation}
The algorithm iteratively refines the estimate of the true image, $J(x, y)$, using the following equation:
\begin{equation}
J_{i+1}(x, y) = J_i(x, y) *[D(x, y) / (J_i(x, y) * P(x, y))]
\end{equation}
Here, $J(x, y)$ is the updated estimate, and the division represents a deconvolution process that takes into account the PSF and the current estimate.

The iterative process continues for a predefined number of iterations or until convergence, progressively reducing the discrepancy between the estimated and observed images. Through this iterative approach, the Richardson-Lucy algorithm effectively restores degraded images, revealing finer details and reducing blurring and noise. Care should be taken when using the Richardson-Lucy algorithm to avoid potential issues with overfitting and the introduction of artifacts.

\subsection{Sample preparation}

Samples were fabricated using lithographic techniques to simulate point sources. The process involved applying a photoresist in the desired pattern onto a 1mm thick slide glass substrate, consisting of a 10 nm TI adhesive layer and a 450 nm Al layer and following a dry etching process. The 450 nm thick Al thin film exhibited a transmittance of less than 0.1 percent at a wavelength of 530 nm, selectively allowing light transmission only through the etched regions of the aluminum layer.

The fabricated sample featured multiple holes with a diameter of 20 $\mathrm{\mu m}$, deliberately chosen for their size to be 3 to 8 times smaller than the diffraction limit of the experimental setup to closely approximate point sources. This choice ensured that the fabricated holes mimicked point sources, facilitating the study and analysis of point source characteristics within the limitations of the experimental setup. 

The sample included three main patterns. The first pattern consisted of a single circular hole primarily utilized for capturing point spread function features. The second pattern comprised two holes spaced 80 $\mathrm{\mu m}$ apart, simulating the presence of a two-point source scenario. Lastly, the third pattern consisted of three equally spaced holes arranged in a triangular array, with a separation distance of 80 $\mathrm{\mu m}$.

\subsection{Point Spread Function Measurement}

The PSF of the experimental imaging system was measured using a single 20 $\mu m$ pinhole target. A Gaussian was fit to the image from the pinhole target for each aperture size, and the result of this fit was used as the known PSF for all algorithms.

\subsection{SNR Calculation}

During the experiment, we ensured a consistent noise level resulting from room lighting. This noise level averaged about 270 photons per pixel, intentionally set to be more than 10 times larger than the combined read and dark noise, which typically falls below 27 photons. To maintain a controlled SNR throughout the measurements, we adjusted the power of the LED signal, while keeping the noise level constant. By controlling the SNR in this manner, we could effectively tune the balance between the signal and noise components in our measurements.
SNR is calculated on a per-pixel basis, with the signal being the mean pixel intensity over 30 frames with the average background subtracted, and the noise being the standard deviation of the pixel intensity over 30 frames. Reported SNRs are pixel SNRs calculated at the ground truth location of the dimmest source in the scene.

\subsection{Quantifying imaging in the sub-Rayleigh limit}

For the evaluation of location estimation accuracy, we employ the root mean squared error ($\mathrm{r.m.s.e.}$). Given the ground truth location $(x_i,y_i)$ and the estimated point $(\hat{x_i},\hat{y_i})$, the $\mathrm{r.m.s.e.}$ is computed as:

\begin{equation}
    \mathrm{r.m.s.e.}= \sqrt{\frac{\sum_i (x_i-\hat{x_i})^2 + (y_i-\hat{y_i})^2  }{N}}.
\end{equation}

To gauge the accuracy of number estimation, we utilize the Jaccard Index \cite{sage2019super}. Each estimated point is classified into one of three categories: True Positive (TP), True Negative (TN), or False Negative (FN). A point is marked as TP if it falls within a distance of $\epsilon$ from the ground truth location, where $\epsilon$ is set at 2.5 pixels in this study. False Positives (FP) are incorrectly estimated points, and FNs represent points that were not detected. The Jaccard Index ($\mathrm{JAC}$) is then calculated as:

\begin{equation}
    \mathrm{JAC}=100 \frac{TP}{TP+FP+FN}.
\end{equation}

To collectively assess the performance of number detection and location estimation, we apply the Jaccard efficiency ($E$) metric \cite{sage2019super}, defined as:

\begin{equation}
    E=100-\sqrt{(100-\mathrm{JAC})^2+\alpha^2 \mathrm{r.m.s.e.}^2}.
\end{equation}

In this analysis, the parameter $\alpha$ which adjusts the weight of each variable, is set at 10.

\section*{Acknowledgements}
This work was partially supported by the Army Research Office (W911NF-21-1-0287) and Defense Advanced Research Projects Agency (DARPA). Partial support also came from Purdue School of Electrical and Computer Engineering. The authors thank Johns Hopkins Applied Physics Laboratory for providing the code for CLEAN, Richardson Lucy Deconvolution, and error metric calculation code. The authors also wish to thank Dr.Fanglin Bao, Dr.Farid Kalhor, and Leif Bauer for discussing the result.

\section*{Data availability}
All data needed are included and discussed in the article. Further code or data is available from the authors upon request.

\section*{Conflict of interest}
The authors declare that they have no conflict of interest.

\section*{References}
\bibliographystyle{unsrt}
\bibliography{SSRI}

\end{document}